| | |
|---|---|
| **Abstract** | When mature *Arabidopsis thaliana* seeds are dormant, their germination is prevented in apparently favourable conditions. This primary dormancy can be released during seed dry storage through a process called after-ripening whose duration can last several months. To reduce this delay, cold atmospheric plasmas (CAP) can be used as sources of reactive oxygen species capable of inducing heterogeneous chemical reactions. While CAP are known to stimulate the germination of various seed species, the relationship between CAP treatments and the amorphous solid state of dry seeds remains unexplored. Here, we demonstrate that seed dormancy can be alleviated using a cold plasma of ambient air and that this alleviation can be amplified for seeds with high water-content (typically 30 %$_{DW}$) or seeds heated at 60 °C during plasma treatment. Differential scanning micro-calorimetry shows that these characteristics control the glassy/rubbery state of the seed cytoplasm. This technique indicates also that a glass transition to the rubbery state strengthens the CAP effects to alleviate seed dormancy. We propose that lower cytoplasmic viscosity can promote the oxidative signaling induced by CAP which, in turn, improves the germination process. |
| **Keywords** | Plasma agriculture, seeds biology, dormancy release, differential scanning calorimetry, *Arabidopsis thaliana*. |

# 1. Introduction

Since the early 21st century, many plasma processes have been developed to improve the quantitative parameters of seed germination such as vigour, homogeneity and germination rate as well as seedlings and plants' growth [1]. Among these processes, two main approaches can be distinguished: (i) the dry plasma approach (also single-step approach) which consists to expose seeds, seedlings or plants directly to a discharge or post-discharge [2] and (ii) the wet plasma approach (also called two-steps approach) which consists to plasma-activate a liquid (water, liquid fertilizer) that is subsequently utilized to imbibe seeds or irrigate plants [3]. A wide variety of seed species has already been studied following these approaches, including oilseed rape, bean, soybean, wheat, tomato, radish, mulungu, lentil [4-12]. Both routes are relatively efficient, even though the dry one seems easier to transfer to industrial applications. Besides, seeds/seedlings treated following the wet approach can only benefit from the indirect effects of plasma, i.e. the long lifespan reactive species present in the liquid medium. On the contrary, the seeds treated following a dry approach benefit all the aforementioned plasma properties (electric field, reactive chemistry, radiation). Among the various devices offering such dry operation, dielectric barrier devices (DBD) appear quite relevant because they can reduce osmotic and saline stresses of seeds like *Arabidopsis thaliana* [13] but also improve the size of crop plants from sowing to harvest [14]. Besides, they can generate a cold atmospheric plasma (CAP) operating in ambient air, hence reducing the investment costs usually associated with low-pressure devices.

Up to this point, regardless of whether a dry or wet plasma method is used, there has been limited research exploring if cold plasma can effectively break seed dormancy. Research led by Gómez-Ramírez et al. has provided evidence that ambient air plasma may indeed address this problem. Their experiment involved treating Quinoa seeds in a DBD at 500 mbar, which resulted in an increase in germination rates from 60% to nearly 100% [15]. Additionally, further studies conducted on seeds of Mimosa Caesalpinia Folia, which suffer from integument physical dormancy, revealed that plasma treatment can boost their germination rates from 6% to 50% [16].

Here, we assess the potential of CAP to release the dormancy of *A. thaliana* seeds, a genus of herbaceous plants from the Brassicaceae family. Dormancy is defined as the inability of seeds to germinate although all the required environmental conditions are apparently satisfied [17]. This adaptation mechanism enables species to survive in their ecosystems, waiting for the optimal season to stimulate seed germination and plant growth [18]. *A. thaliana* seeds are subjected to a physiological non-deep dormancy that blocks their germination after maturation on the mother plants and in which abscisic acid (ABA) and gibberellins (GA) play antagonist roles [19]. Abscisic Acid is the key hormone involved in inducing and maintaining seed dormancy. It helps prevent premature germination during unfavorable conditions, such as dry seasons or cold temperatures. In essence, high levels of ABA within a seed promote dormancy, preventing the seed from germinating until conditions are suitable. Conversely, gibberellins have a promoting effect on seed germination. These key hormones counteract the inhibitory effects of ABA on seed germination. When conditions are appropriate for germination, the levels of gibberellins increase, which stimulates the breakdown of stored food resources within the seed and helps kick-start the germination process [17]. For *A. thaliana*, the highest expression of primary dormancy is usually obtained when dormant seeds are germinated above 20 °C [20].

*A. thaliana* seed dormancy is naturally released during after-ripening which occurs while the dry seeds are stored over a prolonged period, typically for several weeks. The storing conditions, in particular the temperature and the relative humidity, play a crucial role in regulating the breaking of dormancy [21,22]. An interesting assumption is that the combined effect of water content and storing temperature can change intracellular viscosity which, in turn, can release dormancy. In addition to dry after-ripening, other methods can be used on *A. thaliana* seeds, including cold stratification (2-4 days of imbibition at 4°C), light exposure, gibberellin or ethylene treatments [23]. However, they all rely on seed hydration and are therefore concomitant with the germination process itself. Discovering novel methods that can break seed dormancy through a dry route could offer an innovative and intriguing prospect.

From a condensed matter physics perspective, dry seeds can be assimilated to amorphous solid materials owing to their ability to exist either in a glassy state or in a rubbery state depending on their temperature and water content. Hence, after their maturation on the mother plant, orthodox seeds can survive from a severe desiccation by entering into a glassy quiescent state [24]. In such an amorphous glassy state, no enzymatic reaction can theoretically occur except the oxidation or peroxidation reactions related to the production of reactive oxygen species (ROS) [25]. Then, when seed water content (WC) or temperature increases, the cytoplasm can enter into an amorphous rubbery state which presents a lower viscosity [24].

The mechanisms facilitating the release of seed dormancy in a dry (or glassy) state remain largely elusive. Current knowledge indicates that this alleviation happens during after-ripening, when non-enzymatic ROS are generated within the seeds [26]. Then, these reactive species may initiate oxidative signaling during the imbibition phase, leading to alterations in hormone signaling, gene expression, redox regulation and protein oxidation. Although these collective modifications can enhance the germination potential of seeds, it is worth stressing that a balance in ROS levels is always critical [27]. Since excessive accumulation of ROS can impair germination potential and shorten seed viability, ROS concentrations within dry dormant seeds should be maintained within a defined 'oxidative window' to ensure a safe transition to a non-dormant state (where dormancy is released and germination completed) [28]. Delimiting the boundaries of this oxidative window is complicated by the external factors of dry storage (or after-rippening), especially the temperature and the relative humidity which can both influence seed cytoplasmic viscosity. As a result, these two factors can affect seed metabolism and physiology, potentially leading to reduced dormancy alleviation, seed aging and viability loss [29-31]. Despite these insights, a considerable knowledge gap exists regarding how changes in cytoplasmic viscosity impact dormancy release, oxygen diffusion and subsequent non-enzymatic ROS production via autoxidation reactions (within dehydrated seed tissues).

In this work, we hypothesize that both the water content of seeds and the amorphous properties of their tissues may modulate cytoplasmic viscosity and cellular state. These factors could potentially affect oxygen diffusion and ROS generation in seed tissues, subsequently influencing the plasma's role in seed dormancy release. To verify this hypothesis, a CAP of ambient air is generated in a DBD and characterized by electrical measurements, optical emission spectroscopy and mass spectrometry. Then, the amorphous state of the seed – whether glassy or rubbery – is modified and controlled following two different approaches: (i) before the plasma treatment by equilibrating seed water content in atmospheres characterized by specific values of relative humidities, (ii) during the plasma treatment by coupling the DBD with an external gas heating source (heating plate device). The effects of CAP as well as its efficiency on seed dormancy release are assessed under these conditions, while glass transitions are measured by differential scanning micro-calorimetry (DSC) which provides information on cytoplasmic glass transition at cell level.

## 2. Material & Methods

### 2.1. Plant material and germination assay

*Arabidopsis thaliana*, ecotype Col-0, seeds are grown and harvested dormant, following the method described by *Leymarie et al.* [20]. Two independent seed batches are used in this study. For germination assays, 4 replicates of 100 seeds each are distributed on a layer of cotton wool moistened with demineralized water in glass Petri dishes (9 cm in diameter) and placed at 25 °C in darkness, conditions in which *A. thaliana* seed primary dormancy is strongly expressed [20]. Seed germination is scored twice a day for approximately 12 days. A seed is considered as germinated once the radicle protrudes the envelope. Each germination curve is obtained by monitoring samples of 400 seeds, with 100 seeds per Petri dish so that the mean values are calculated on 4x100 seeds with standard deviation.

### 2.2. Influence of water content and temperature on the dormancy release of *Arabidopsis thaliana* seeds

Before studying the effects of cold atmospheric plasma on dormancy release, one has to verify whether exogenous factors such as seed water content and temperature can induce beneficial/detrimental effects. This is important to understand whether combining cold atmospheric plasma with these factors are likely to induce synergetic effects to release dormancy.

To change and control seed water content, the following procedure is followed. First, seeds are placed in tightly closed jars at 20 °C and equilibrated for 4 days at various WC over silica gel or over saturated saline solutions. Hence, the relative humidity of these atmospheres can be changed and controlled, as reported in **Table 1** and according to the method described by *Vertucci et al.* [32]. Then, seed water content (WC) is expressed on a dry weight basis expressed as $\%_{DW}$ (or $g_{H_2O} \cdot g_{dw-1}$) which is determined after oven drying for 24 h at 105 °C. As shown in **Fig. 1a**, increasing the water content from $1\%_{DW}$ to $30\%_{DW}$ increases the germination rate from roughly 1% to 20%.

As an alternative to changing seed water content, it is possible to keep native water content of *A. thaliana* seeds while modifying their temperature using a hot plate. Hence, **Fig. 1b** shows that the germination rate can be increased from approximately 1% to 13%, which is low but non negligible.

| Medium of equilibration | | RH (%) | WC range (%$_{DW}$) | Equilibration duration |
|---|---|---|---|---|
| Silica gel | | - | 2.5 | 1.3 |
| Saline solutions | $ZnCl_2$ | 9.5 | 2.9 | 4 days |
| | LiCl | 17.0 | 4.8 | |
| | $CaCl_2$ | 28.5 | 5.2 | |
| | $MgCl_2$ | 40.0 | 6.3 | |
| | $Ca(NO_3)_2$ | 56.5 | 7.5 | |
| | NaCl | 75.0 | 11.3 | |
| | KCl | 83.5 | 15.1 | |
| | $KH_2PO_4$ | 93.0 | 31.0 | |

*Table 1. Controlling water content of Arabidopsis thaliana seeds by confining them in atmospheres with different relative humidities.*

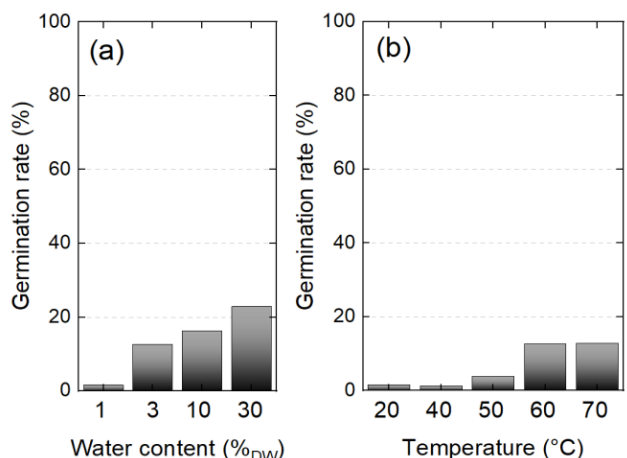

*Figure 1. Effects of exogenous factors on the germination rate of A. thaliana seeds (a) Influence of seed water content at 20 °C, (b) Influence of seed temperature with WC = 3%$_{DW}$.*

## 2.3. Plasma process

As sketched in **Fig. 2**, seeds are treated using a dielectric barrier device (DBD) in a plan-to-plan configuration with a gap of 1 mm, i.e. 3 times larger than the average diameter of *A. thaliana* seeds. The DBD has two electrodes: (i) an electrode biased to high voltage, 30x40 cm² in area, composed of a stainless-steel mesh with a 100 µm nominal opening, a wire diameter of 71 µm and an open area of 34 % and (ii) a counter-electrode which is grounded, 30x40 cm² in area, bulk alumina plate. The dielectric barrier is a 2mm thick glass plate on which the mesh electrode is fixed. Optionally, a heating plate device (Froilabo EM130 Hotplate Magnetic Stirrer) is in contact with the counter-electrode to control temperature in the gap of the DBD between 30 °C and 80 °C with an accuracy of ± 2 °C.

The mesh electrode is powered with a high voltage generator composed of a function generator (ELC Annecy France, GF467AF) and a power amplifier (Crest Audio, 5500W, CC5500). The power supply is electrically protected by a ballast resistor (250 kΩ, 600 W) presented in **Fig. 2**. The peak amplitude of the voltage applied on the mesh electrode reached 7 kV at 145 Hz. Typically, 1000 seeds of *A. thaliana* can be placed on the grounded electrode, occupying a volume as low as $V_{seeds}$ = 1000 x $V_{seed}$ = 14 mm³. Since the gap volume is $V_{GAP}$ = 30 x 40 x 0,1 cm³ = 120 cm³, the volumetric filling rate of the device is only 0.01 %.

## 2.4. Electrical characterization

Voltages and currents of the DBD are measured using a digital oscilloscope (model wavesurfer 3054 from Teledyne LeCroy company), with a 500 MHz bandpass and a 4GS/s sample rate. The voltage applied to the mesh electrode is measured with high voltage probes (Tektronix P6015A 1000:1, Teledyne LeCroy PPE 20 kV 1000:1, Teledyne LeCroy PP020 10). The discharge current is monitored using a current transformer placed between the metal plate counter-electrode and the ground and enables to rebuild the distribution of plasma current peaks. A measurement capacitor ($C_m$ = 80 nF) is inserted between the counter-electrode and the ground to measure the voltage ($V_{DBD}$) and the charge ($Q_{DBD}$) of DBD. Then, Lissajous curves are plotted to measure the energy dissipated in the DBD. Multiplying this value by the operating frequency provides the electrical power injected in the DBD.

## 2.5. Optical emission spectroscopy (OES)

Optical emission spectroscopy is achieved to identify the radiative species emitted by plasma. The spectrometer (Andor, SR-750-B1-R) is used in the Czerny Turner configuration with a focal length of 750 mm over a 200-850 nm spectral range. The device includes an optical fiber (Leoni fiber optics SR-OPT-8014, 100 µm diameter), a diffraction grating (1200 grooves.mm$^{-1}$) blazed at 500 nm and an ICCD camera (Andor Istar, 2048 × 512 imaging array of 13.5 µm × 13.5 µm pixels). To collect a maximum of plasma emission with a 1 mm spatial resolution, a converging lens (ThorLabs, LA4380-UV, f = 100mm) is placed between plasma and the optical fiber.

## 2.6. Mass spectrometry

Mass spectrometry is performed to identify the relevant species from the plasma phase. The mass spectrometer used in this study is a quadrupole-based mass spectrometer (Model HPR-20 from Hiden Analytical Ltd). Air plasma chemical species are collected by a quartz capillary whose inlet is fixed onto a 2-axis stages plate placed in the middle-left of the plasma reactor (see **Fig. 2**). The 1 m long capillary is flexible, chemically inert and heated at 200 °C to prevent chemisorption. Then, a three-stage differentially pumped inlet system separated by aligned skimmer cones and turbo molecular pump, enables a pressure gradient from 10⁵ bar to 10⁻⁷ bar at the entrance of the 70-eV energy ionization chamber. The species detected by mass spectrometry are quantified by measuring their intensities when the gas is ionized (plasma) and then subtracting the corresponding intensities when the gas is non-ionized (background). In both cases, the measurements are performed in steady state, obtained once the pressure is stabilized after the ignition of the filament (about 1h).

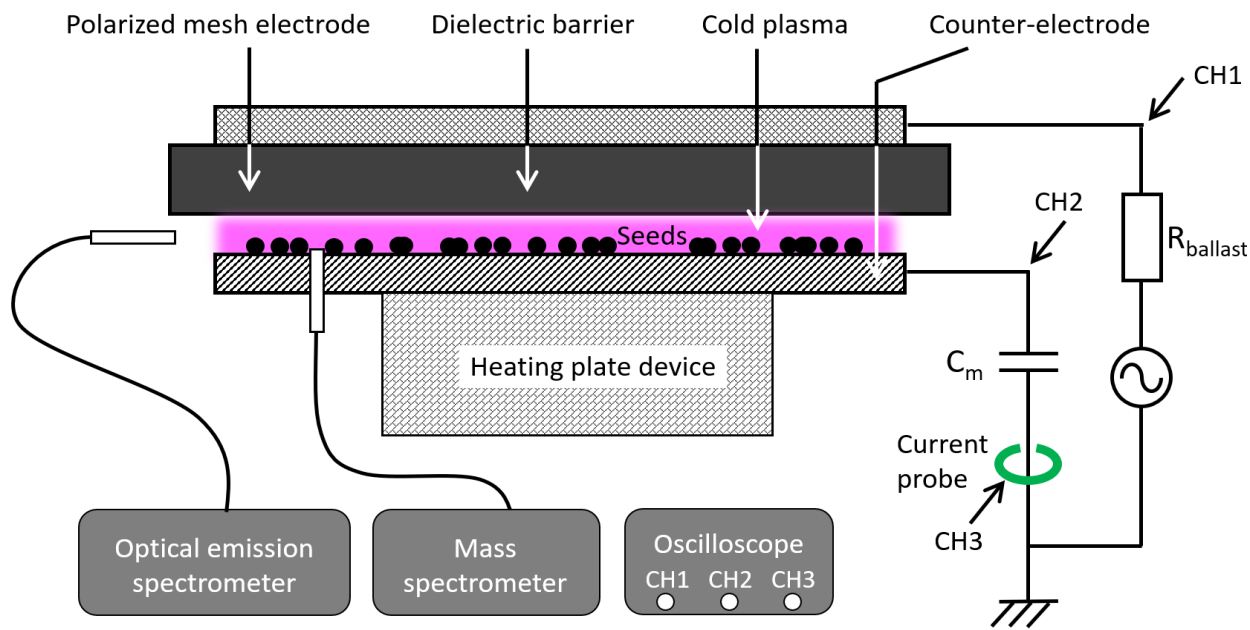

*Figure 2. Sketch diagram of the DBD utilized to treat seeds by ambient air cold plasma.*

## 2.7. Differential scanning micro-calorimetry (DSC)

A differential scanning micro-calorimeter (DSC 30, Mettler Toledo) is used to measure enthalpy variations of the seeds and deduce the glass to rubber transition temperature (i.e. glass transition temperature, $T_g$). All the DSC measurements are carried out on samples composed of approximately 20 mg of *A. thaliana* seeds. Each sample is placed into a 40 μL aluminium crucible which is hermetically sealed to suppress the endothermic evaporation, vaporization or sublimation of volatile substances in the furnace. The seeds are exposed to a 10 °C/min ramp between 0 °C and 100 °C, so that the glassy-rubbery transition is highlighted by a discontinuity of the seeds sample thermal capacity upon the 10 °C/min rate. The $T_g$ is estimated as the mid-point of the temperature range associated with the shift in specific heat. Between 8 and 12 DSC scans are performed with seeds equilibrated at each relative humidity used in this study.

## 2.8. Statistical analyses

All the statistical analyses are performed using OriginLab Software statistical tools, including curve fitting, Mann-Whitney and ANOVA tests to determine significant differences between control group and plasma group on the total germination rates.

## 3. Results

### 3.1. CAP process strongly depends on the applied frequency and external gas heating source

#### 3.1.1. The plasma spreading area is strongly influenced by the applied frequency

Since the volumetric filling rate of the device is only 0.01 %, the action of the seeds on the plasma electrical properties can be neglected. As shown in **Fig. 3a**, the voltage applied to the DBD has a sine waveform, 7 kV in magnitude, while the current intensity shows two components: (i) a dielectric component proportional to the derivative of the applied voltage and (ii) a conduction component composed of current peaks (with typical widths of a few tens of nanoseconds) that represent the micro-discharges randomly distributed over a time window delimited by the breakdown and extinction voltages. At first sight, the spatial distribution of these micro-discharges is uniform in the DBD, hence ensuring a homogenous treatment of the seeds. However, owing to the large size of the electrodes ($S_{elec}$ = 30 x 40 cm$^2$), the plasma spreading area does not match necessarily to $S_{elec}$, especially because of the frequency value. As indicated in **Fig. 3b**, while the distribution of the micro-discharges is homogeneous over the entire electrode surface for f < 200 Hz, its spreading area gradually decreases for higher frequencies, leaving dead zones where the plasma never forms throughout processing. This phenomenon can be explained by considering that on each period, the same global electric charge Q is divided into a certain number of smaller charges q, each one being associated to a current peak and thus to a micro-discharge. According to **Fig. 3a**, the current peaks are randomly distributed over a fraction of the period (active region). If the signal frequency is increased, the duration of this active region is reduced. Since the same charge Q must still be delivered over this shorter active region, the current peaks become less numerous but reach larger amplitudes. This temporal distribution is obviously correlated to a spatial distribution of the micro-discharges which are also becoming rarefied at higher frequencies, as illustrated in **Fig. 3b**.

To achieve the plasma power measurements, a capacitor ($C_m$) is introduced in the experimental setup. To ensure that this component will be non-intrusive, its capacitance must be at least 10 times higher than that of the dielectric barrier, estimated here at 3.7 nF. The **Fig. 3b** represents the electrical power deposited in the DBD as a function of the applied frequency for different values of $C_m$. It turns out that a maximum power (18 W) is deposited for f = 145 Hz and $C_m$ = 80 nF: two values that are then maintained in all the subsequent experiments.

### 3.1.2. Electrical and chemical properties of CAP are modified by the external gas heating source

Seed cytoplasmic glass transition depends on seed temperature. To better understand how CAP combined with seed thermalization can synergistically improve dormancy release, first it is important to study the influence of the external heating source on the plasma electrical properties, and more specifically on the current peaks, as clearly highlighted from **Fig. 4a** to **Fig. 4e**, considering values of 25 °C, 40 °C, 50 °C, 60 °C and 70 °C. A rise in temperature simultaneously leads to an increase in the amplitude of the peaks and a decrease in their number, resulting in a higher charge-per-filament. Knowing the area covered by the micro-discharges within the DBD, a surface charge density can be deduced, as reported in **Fig. 4f**. For temperatures lower or close to 50 °C, the surface charge density is always lower than 0.05 µC.cm$^{-2}$ whereas its value drastically increases to 0.25 µC.cm$^{-2}$ at 60 °C and 0.32 µC.cm$^{-2}$ at 70 °C. To verify the influence of the temperature on the electrical power, the charge $Q_{DBD}$ is plotted as a function of the voltage measured in the DBD in **Fig. 4g**. Each resulting Lissajous figure looks like a parallelogram whose upper and lower edges correspond to the capacitance of the dielectric barrier ($C_{db}$) while the left and right edges correspond to $C_{db}$ in series with the gas capacitance ($C_g$). The area covered by each Lissajous parallelogram corresponds to the electrical energy from which the electrical power is deduced by multiplying the energy by the frequency. Surprisingly and whatever the temperature, the electrical power is quite constant with a value close to 20 W.

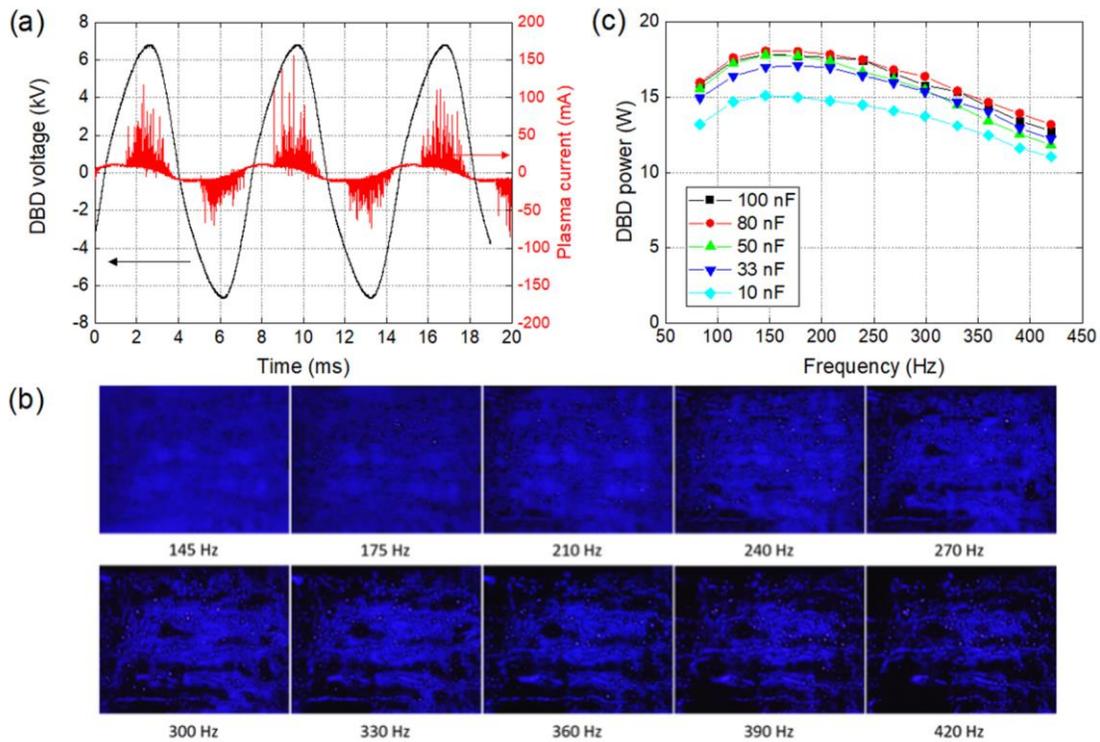

*Figure 3. Electrical characterization of ambient air cold plasma generated in DBD. (a) Typical waveform of the plasma voltage and current for $C_m$ = 80 nF, f = 145 Hz. (b) DBD electrical power versus frequency for different measurement capacitors. (c) Images of the plasma observed through the HV mesh electrode for different frequencies at $C_m$ = 80 nF.*

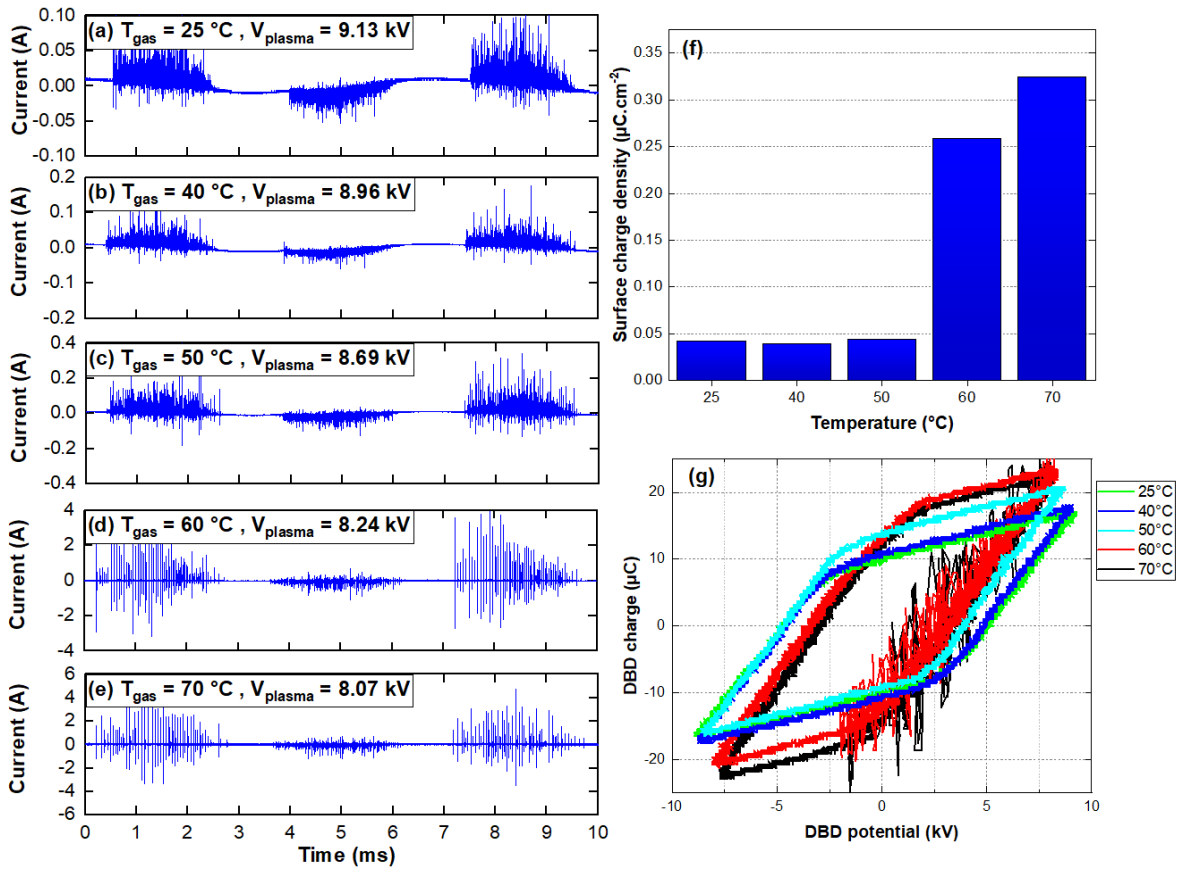

*Figure 4. Electrical characterization of the DBD as a function of temperature. Waveforms of discharge current measured in the DBD (80 nF, 145 Hz) (a) at 25 °C, (b) 40 °C, (c) 50 °C, (d) 60 °C, (e) 70 °C. (f) Surface charge density estimated through 9 periods vs gas temperature. (g) Lissajous figures.*

Since the external gas temperature can change the electrical properties of CAP, it is important to check whether the heterogeneous chemical properties of CAP are also influenced, in particular through the production of active species. **Fig. 5a** presents the optical emission spectrum of the ambient air cold plasma generated in the DBD. The emission lines and bands are part of the UV-VIS range (200-800 nm) and are derived from the energetic processes producing reactive oxygen and nitrogen species. If the vibrational bands of the Second Positive System of molecular nitrogen ($C^3\Pi_u - B^3\Pi_g$) are clearly observed [33-35], it is worth notifying the absence of oxygen line at 777 nm. This can be explained by the fact that in ambient air cold plasma, the O ($^5P$) and O ($^3P$) lines are quenched by $N_2$ and $O_2$ before being detected by OES [33]. A large part of atomic oxygen could also be involved in fast kinetic reactions, i.e. recombination into molecular oxygen or impinging with molecular oxygen to form ozone. Interestingly, emission bands of NO are detected [36] in the 220-260 nm region while no emissive band of OH radicals is evidenced at 309 nm. Finally, just as the temperature influences the electrical properties of the plasma, it influences its optical properties as well. As observed in **Fig. 5b**, the maximum emission of the $N_2$ band at 337 nm increases with temperature in a range between 25 and 70 °C. Indeed, not only the temperature of the gas increases, but also the temperature of the device, in particular the dielectric barrier whose permittivity depends on the temperature.

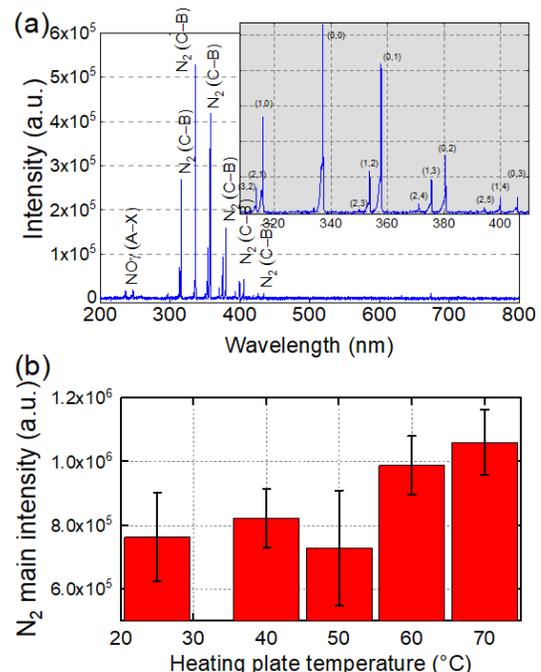

*Figure 5. (a) Optical emission spectrum of ambient air cold plasma and (b) $N_2$ main lines intensities as a function of the applied temperature (7 kV, 145 Hz, 1 mm gap).*

In addition to the gas phase analysis carried out by optical emission spectroscopy, mass spectrometry is utilized to measure active species, whether emissive or not and considering two cases: DBD alone (gas temperature = 25 °C) and DBD combined with heating plate device (gas temperature = 60 °C). Each species is measured by plotting the ratio of its intensity by the air intensity (corresponding roughly to 78 % of $N_2$ intensity and 22 % of $O_2$ intensity). As shown in **Fig. 6**, CAP at 25 °C is composed of non-reactive stable molecules ($N_2$, $O_2$, $H_2O$) as well as reactive species of short lifespan (O, NO and $NO_2$) and long lifespan ($O_3$). Although it is common to measure OH radicals in air plasmas, their amount remains here undetectable, consistently with the optical emission spectroscopy results (**Fig. 5a**). Unsurprisingly, the external gas heating source thermalized at 60 °C has no significant influence on the $N_2$ and $O_2$ components measured in CAP but leads to a strong decrease in the water vapour component. Interestingly, the NO and $NO_2$ reactive species are produced in larger amounts and could explain the beneficial effects of the CAP/heating plate combination. Finally, the variations in atomic oxygen (Z/q= 16) and ozone (Z/q=48) are not significant. Therefore, it is possible to deduce that plasma-triggered dormancy alleviation is strongly related to the production of O, NO, $NO_2$ and $O_3$ reactive species and could be even more enhanced by a higher production rate of NO and $NO_2$ species.

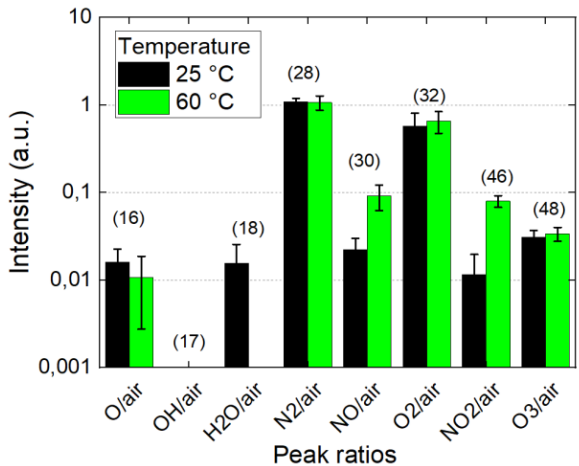

*Figure 6. Intensity of peak ratios measured by mass spectrometry in CAP (air, 7 kV, 145 Hz, 1 mm gap) combined or not with the external gas heating soruce, i.e. at 25 °C or 60 °C. The Z/q values of the detected species are indicated in parenthesis.*

## 3.2. Seed dormancy can be released by CAP and enhanced by controlling seed material properties

### 3.2.1. State diagram of *A. thaliana* seeds highlights glass transition temperature, rubbery and glassy states

Differential scanning micro-calorimetry (DSC) has been achieved on the *A. thaliana* seeds to determine their glass transition temperature ($T_g$) and therefore whether their cytoplasm is in glassy or rubbery state. First, DSC measurements are achieved on seeds to plot their heat flow versus a ramp of temperature between 30 and 75 °C, as shown in **Fig. 7a**. The glass transition temperature is characterized by a second-order transition, i.e. a shift in the baseline of the DSC thermogram, here estimated at 51.47 °C. Then, multiple DSC measurements are carried out on seeds equilibrated at various WC (ranging from 1.7 to 31.0 %$_{DW}$) to determine the corresponding values of $T_g$. Finally, these glass-to-rubber transition temperatures are reported versus seed water content, as shown in **Fig. 7b**, also called a state diagram. Within the WC range, $T_g$ decreases from 61 °C to 20 °C following an exponential decay whose fitting equation is given by $T_g = 64.8\, e^{-3.7 \times WC}$. The accuracy of the fit is validated by the R-square coefficient as high as 0.95. Therefore, the state diagram permits to unambiguously delineate glassy and rubbery state regions versus both temperature and WC in *A. thaliana* seeds.

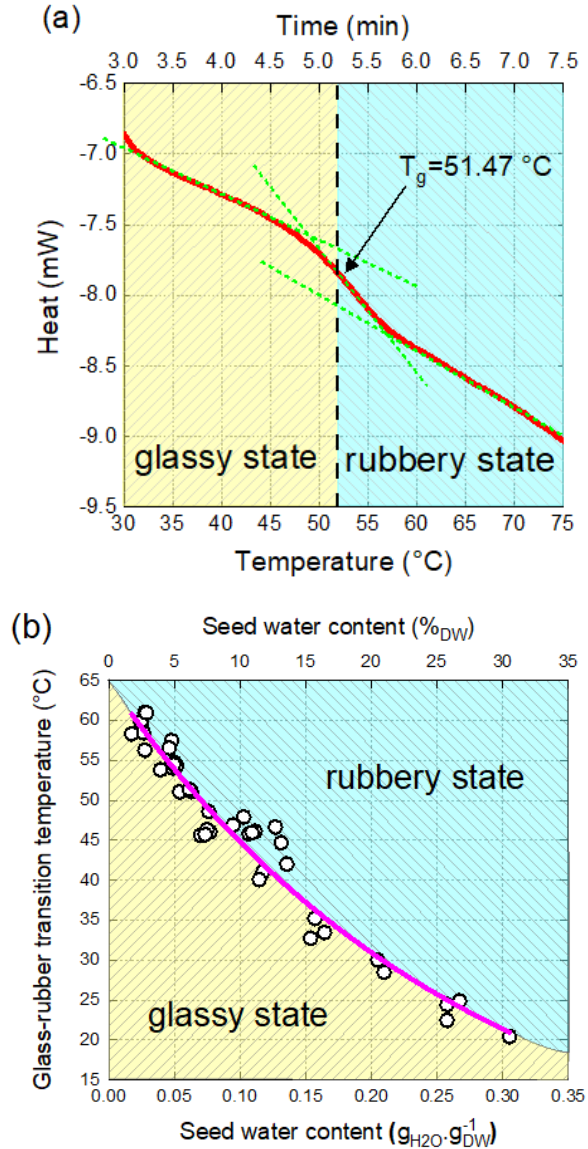

*Figure 7. Seed glass transition characterization in Arabidopsis thaliana seeds. (a) a DSC thermogram of 20 mg Arabidopsis seeds of WC = 6.5 %$_{DW}$ and (b) state diagram of Arabidopsis thaliana seeds showing glass transition temperature (purple line) vs WC; each white circle being an experimental data extracted from a DSC thermogram.*

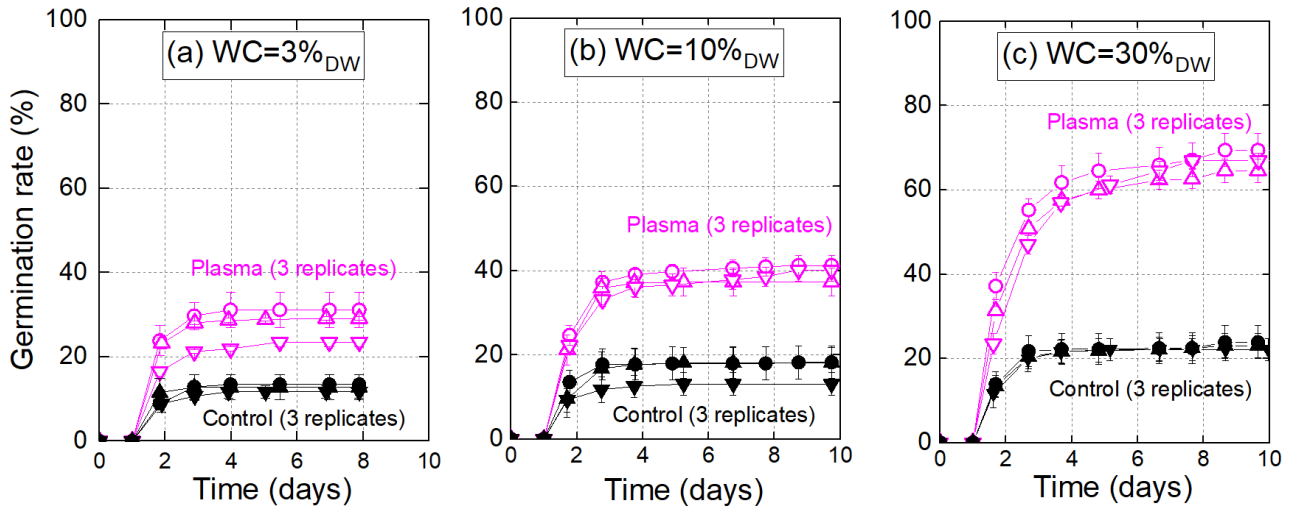

*Figure 8. Germination at 25 °C of seeds equilibrated for 4 days at (a) 1 %$_{DW}$, (b) 10 %$_{DW}$ and (c) 30 %$_{DW}$ water contents (WC, dry weight basis) and exposed to CAP (plasma) for 15 min or not (control). For each replicate, means +/- SD correspond to 400 seeds dispatched into 4 Petri dishes with 100 seeds per Petri dish.*

### 3.2.2. Increasing seed WC to reach the rubbery state, improves the plasma-triggered dormancy release

If CAP can unambiguously release seed dormancy, it is important to decipher whether this effect can be amplified depending on seed WC and therefore on seed cytoplasmic glass transition. To verify this hypothesis, we consider freshly harvested dormant seeds that we equilibrate in water following protocol introduced in section 2.2. In this work, the three following water contents are investigated: 1 %$_{DW}$ (lower value which corresponds to seeds in anhydrous state), 10 %$_{DW}$ (intermediate value which corresponds to the seeds when freshly harvested) and 30 %$_{DW}$ (higher value which corresponds to the mor elevated moisture a seed can contain without directly triggering germination). Considering that the plasma gas temperature is around 30 °C, the state diagram in **Fig. 7b** shows that the seeds of the two first groups are in a glassy state while those from the third group are in a rubbery state. Each group is placed in the DBD where an air plasma is generated at V = 7 kV, f = 145 Hz for 15 min. Germination monitoring of the seeds - whether treated or untreated - is reported in **Fig. 8a**, **8b** and **8c** for the low, medium and high WC conditions, respectively. While the untreated seeds are dormant and exhibit germination rates that never exceed 20 % (whatever the replicate), all the seeds exposed to CAP show increased germination rates with a clearly observed trend: the higher the initial seed WC, the higher the final germination percentage. Indeed, germination rate increases from 10 % to 30 % for seeds in a glassy state (WC = 1 %$_{DW}$) and to a value as high as 70 % for seeds in the rubbery state (WC = 30 %$_{DW}$).

The importance of seed water content on the plasma-enhanced dormancy release is clearly highlighted in **Fig. 9**. Mann-Whitney tests have been performed between control and plasma groups for each water content condition to characterize significant statistical differences. It is found that for the three WC groups, the control and the plasma distributions are significantly different (p-values < 0.0001).

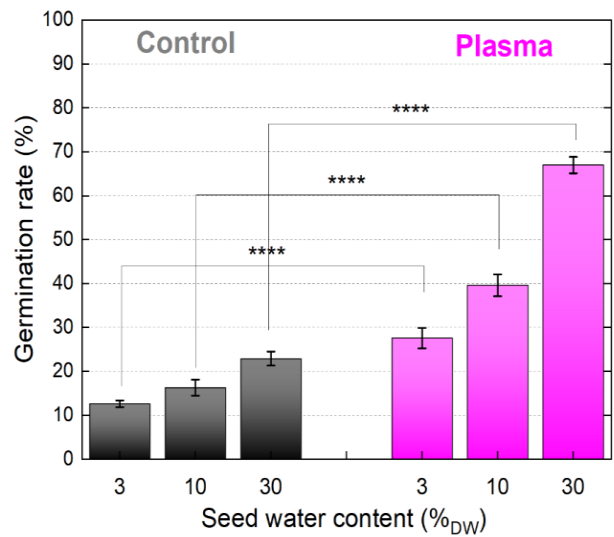

*Figure 9. Germination rates of Arabidopsis thaliana seeds measured 10 days after imbibition and stored at 25 °C. Before imbibition, seeds are equilibrated for 4 days in three different environments to obtain water contents of 1 %$_{DW}$, 10 %$_{DW}$ and 30 %$_{DW}$ (DW: dry weight basis). Three groups of seeds are untreated (Control) while three others are exposed to CAP (plasma) for 15 min. Means +/- SD of 3x4 replicates (****: p value < 0.0001).*

### 3.2.3. Increasing external temperature upon CAP treatment to reach the rubbery state, improves dormancy release

We have demonstrated that CAP alleviates dormancy more efficiently if the seeds are in the rubbery state, obtained by increasing seed WC without changing external temperature. To reinforce this result, and therefore confirm the importance of treating the seeds in this specific amorphous state, we propose an alternative approach that relies on the state diagram in **Fig. 7b**. This figure predicts that the seed rubber state can be obtained at constant WC, for example as low as 7 %$_{DW}$, by increasing temperature higher than 50 °C. Since plasma gas temperature can never exceed 25 °C even after 60 min of operation, the DBD is coupled with a heating plate device, as depicted in **Fig. 2**, and which behaves as an external gas heating source. For each of the experiments described below, the external gas heating source is switched on 30-45 min before achieving the plasma treatment so as to reach a permanent regime, i.e. without significant temperature variation (< 2 °C). Once the DBD is thermalized to the set temperature, the seeds are placed in the interelectrode gap. Temperature of DBD and seeds is monitored in real time using a laser beam thermometer.

First, the effect of the plasma exposure duration is studied on seed germination at 20 °C (**Fig. 10a**). Fifteen, 30 and 60 min of treatment are performed using seeds at 7.0 %$_{DW}$. At this WC value, glass transition occurs at a temperature slightly higher than 50 °C (**Fig. 7b**). In this experiment, the temperature is initially 20 °C (± 1 °C) and gradually increases until reaching a maximum value of 30 °C after 60 min of treatment, which nevertheless remains lower than $T_g$ (**Fig. 7b**). The longest treatment increases germination from 2 % to 50 % while the 15 and 30 min treatments have only a slight beneficial effect on seed dormancy release (**Fig. 10a**).

In order to assess a potential synergy resulting from plasma treatment and thermally-controlled glass transition, another experiment is achieved where two groups of seeds at 7.0 %$_{DW}$ are considered: (i) seeds placed in the DBD during 20 min while plasma is off and temperature increasing from 40 to 70 °C and (ii) seeds placed in the DBD where they are heated for 5 min when plasma is off and then 15 min when plasma is on. As shown in **Fig. 10b**, the heating procedure alone can stimulate seed germination but its value never exceeds 15 %, which suggests that thermal effects alone do not release seed dormancy. Conversely, the 15 min plasma treatments alleviate drastically seed dormancy for temperatures higher than 50 °C, e.g. a germination rate as high as 48 % is obtained at 60 °C (versus only 2 % at room temperature). Combining CAP and heating has spectacular synergistic effects on dormancy alleviation although this operates on a thermal window that remains limited. Indeed, the germination rate obtained at 70 °C (36 %) is lower than the one obtained at 60 °C (48 %).

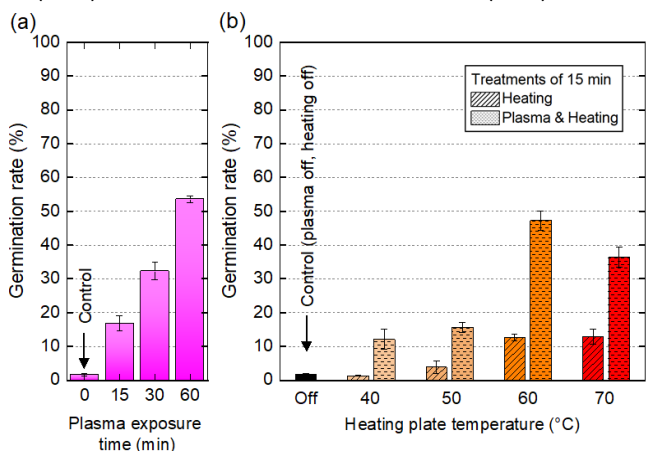

*Figure 10. Germination rates of A. thaliana seeds (WC = 7 %$_{DW}$) measured 10 days after imbibition and stored at 25 °C. (a) Influence of CAP exposure time at room temperature, (b) Influence of CAP for 15 min combined or not with heating at different temperatures. Means +/- SD of 4x4 replicates.*

# 4. Discussion

## 4.1. CAP generated in ambient air as a tool to break physiological seed dormancy

We demonstrate that exposing primary dormant *A. thaliana* seeds for only 15 min to an ambient air cold plasma can lead to a significant alleviation of their dormancy. These effects are obtained by maximizing the plasma power with f = 145 Hz and $C_m$ = 80 nF, as shown in **Fig. 3**. Besides, the electrical behaviour of CAP completely changes by increasing the temperature of the DBD. As shown in **Fig. 4a**, we observe a shift of electrical properties above 50 °C which remains misunderstood. The current waveforms are clearly modified from 60 °C; the intensity of the current peaks increases radically until values of Ampere as well as the surface charge density that is multiplied by a factor of 7 (**Fig. 4b**). Increasing the temperature reduces the relative humidity in the interelectrode gap and could explain variations in terms of current peak amplitudes and time spreading of the micro-discharges [37]. The dielectric barrier is also affected by the rise of temperature as evidenced in **Fig. 4g**. The change in the slope of the upper/lower edges of the Lissajous curves indicate a slight increase in $C_{db}$ due to temperature. This may result from an increase in the relative permittivity of the glass state, as already evidenced in other experiments [38]. Mass spectrometry also attests the increase in NO and $NO_2$ intensity (**Fig. 6**) when increasing the gaseous temperature of CAP.

Our results demonstrate that CAP can efficiently release dormancy when seeds are in a rubbery state, obtained here by changing the seed water content or the temperature during the treatment. Generally, the effect of CAP on seed dormancy is limited to physical dormancy with integument or coat features that limit the water uptake and therefore reduces the germination rate [15,16]. Interestingly, a study has also demonstrated how CAP can break physiological dormancy of radish seeds regarding their storage time after harvest [39]. It is clearly demonstrated that CAP generated in ambient air by DBD can improve the germination rate after different storage times, by measuring the change in the balance of ABA and GA phytohormones after plasma treatment. In our study, we significantly improve the germination rate of primary dormant *A. thaliana* seeds without changing time of storage but by playing on the seeds' amorphous solid state, hence proving that CAP is an appropriate method to release dormancy associated with the seed glass transition.

## 4.2. Glass transition temperature is a pinpoint for releasing seed dormancy by CAP

One of the major findings of our work is to demonstrate that CAP can release seed dormancy and that the efficiency of this effect depends upon intrinsic physicochemical properties of cytoplasm. This finding requires establishing a state diagram of *A. thaliana* seeds which, to our knowledge, has never been performed so far. The state diagram in **Fig. 7b** shows that for a seed WC increasing from 0.017 to 0.310 $g_{H_2O}.g^{-1}$, $T_g$ monotonically decreases from 60 °C to 20 °C because water is a plasticizer of glasses that loosen molecular connections within the glassy matrix [40]. Such trends have also been observed for bean seeds [41] and sunflower seeds [42].

*August et al., J. Phys. D: Appl. Phys., Vol. 56, 415202, 12 pages (2023), https://www.doi.org/10.1088/1361-6463/ace36e*

Here we have taken advantage of the knowledge provided by the state diagram of *A. thaliana* seeds: by changing seed WC or process temperature, it is indeed possible to control the glassy/rubbery state of the seed cytoplasm, and therefore improve the CAP effects in releasing dormancy. We show that whatever the means to ensure glass transition from a glassy to a rubbery state, CAP becomes very effective when cytoplasm is in a rubbery state. Practically, this is reached when seed WC is 30 %$_{DW}$ at ambient temperature (~25 °C) or when seed temperature is 60 °C at 7 %$_{DW}$ (see state diagram **Fig. 7b**). This indicates that the reduction of cellular viscosity promotes CAP effect on seed dormancy release. Glasses are known to dramatically restrict molecular mobility and therefore a wide range of enzymatic and non-enzymatic reactions, and this status is generally associated with seed survival in the dry state [23]. However, in *Pinus densiflora* seeds, gaseous oxygen has been demonstrated to diffuse more easily in glassy state than in rubbery state and to participate in oxidative reactions leading to seed ageing [43]. The shift to a rubbery state that increases the beneficial effect of plasmas on dormancy release, suggests that efficient CAP treatments require some cytoplasmic mobility. Interestingly, the fact that CAP is efficient both when increasing seed WC to 30 %$_{DW}$ and when increasing temperature to 50 °C of dry seeds allow to rule-out that the effect of plasmas strictly depends on an active metabolism. At 30 %$_{DW}$, indeed, seeds contain free water which allows metabolism, including respiration, but at 7 %$_{DW}$ (seed moisture content when seeds are treated at 60 °C) seeds contain only bound water that is not available for metabolic activity, as described by water sorption isotherms [24]. Since it is known that seed dormancy release is controlled by oxidative processes [27], we therefore postulate that seed response to CAP is enhanced owing to a better diffusion of the plasma-generated active species (especially ROS and RNS), when the seed tissues are in the rubbery state. For remind, alleviation of seed dormancy in the dry state can be explained by a selective oxidation of negative regulators of seed germination that results from auto-oxidative reactions and ROS generation [26,27]. Further studies could be achieved to investigate whether CAP treatments can trigger ROS generation within dry seeds.

## 5. Conclusion

In this experimental work, we have demonstrated that CAP generated in ambient air can release the dormancy of *A. thaliana* seeds and that the response of the seeds to such treatment can be improved by placing them in rubbery rather than in glassy state. Such glass transition is expected to increase cytoplasmic mobility of the seeds at cell level, i.e. reduce cytoplasmic viscosity.

Considering the state diagram of *A. thaliana* seeds through DSC measurements, we have demonstrated that the rubbery state can be obtained either by increasing seed water content to 30 %$_{DW}$ before CAP treatment or by increasing the gas temperature up to 60 °C during the CAP treatment. This type of amorphous state is characterized by lower cytoplasmic viscosity that is therefore expected to be responsible for enhancing CAP effects to release *A. thaliana* seed dormancy. Our findings should open novel areas of CAP utilization to rapidly release seed dormancy of crop species.

**Supplementary Materials:** The following supporting information can be downloaded at: www.mdpi.com/article/10.3390/plants11202694/s1, Table S1: Experimental primer sequence.
**Author Contributions:** C.B. and T.D. designed the research; J.A. performed research; J.A., T. D. and C.B. analyzed the data; J.A. wrote the paper; and C.B., T.D. supervised the project. All authors have read and agreed to the published version of the manuscript.
**Funding:** This work was supported by a PhD grant from Sorbonne Université and received financial state aid as part of the $^{PF2}$ABIOMEDE platform co-funded by « Région Ile-de-France » (Sesame, Ref. 16016309) and Sorbonne Université (technological platforms funding). The authors express their gratitude to the Institute of Environmental Transition and the Physics Department of Sorbonne University for the financial support provided to their research within the framework of the AMI project.
**Informed Consent Statement:** Informed consent was obtained from all subjects involved in the study.
**Data Availability Statement:** The data that support the findings of this study are available upon reasonable request from the authors.
**Conflicts of Interest:** The authors declare no conflict of interest.

## 6. References

[1] Attri P et al. 2020 *Processes* **8**(8), 1002 https://doi.org/10.3390/pr8081002
[2] Dufour T et al. 2021 *J. Phys. D : Appl. Phys.* **54**(50) 505202, https://doi.org/10.1088/1361-6463/ac25af
[3] Guo D et al. 2021 *Sci. of food and agriculture*, **101**(12), 4891-4899 https://doi.org/10.1002/jsfa.11258
[4] Ling L *et al.* 2015 *Sci. Rep.* **5** 13033 10.1038/srep13033
[5] Bormashenko E *et al.* 2012 *Sci Rep* **2,** 741 10.1038/srep00741
[6] Zahoranova A *et al* 2016 *Plasma Chem. Plasma Process.* **36** 397–414. 10.1007/s11090-015-9684-z
[7] Zhou R *et al* 2016 *Sci. Rep.* **6** 32603 doi: 10.1038/srep32603
[8] Jiang J *et al* 2014 *PLoS One* **9** e97753 10.1371/journal.pone.0097753
[9] Lu Q *et al* 2014 *Plasma Processes Polym.* **11** 1028–1036 10.1002/ppap.201400070
[10] Kitazaki S *et al* 2014 *Curr. Appl. Phys.* **14**, S149–S153 10.1016/j.cap.2013.11.056
[11] Alves Junior C *et al* 2016 *Sci. Rep.* **6** 33722 10.1038/srep33722 (2016).
[12] Zhang S, Rousseau A and Dufour T 2017 *RSC Adv.* **7** 31244-31251 10.1039/C7RA04663D
[13] Bafoil M *et al.* 2019 *Sci Rep* **9** 8649 10.1038/s41598-019-44927-4

[14] Koga K et al. 2016 *Appl. Phys. Express.* **9** 016201 10.7567/APEX.9.016201
[15] Gómez-Ramírez A *et al* 2017 *Sci. Rep.* **7** 5924 10.1038/s41598-017-06164-5.
[16] da Silva ARM *et al* 2017 *Coll. Surf. B: Biointerfaces* **157** 280–285 10.1016/j.colsurfb.2017.05.063
[17] Bewley JD 1997 *The Plant Cell* **9** 1055-1066 10.1105/tpc.9.7.1055
[18] Baskin CC and Baskin JM 1998 *Seeds: ecology, biogeography, and evolution of dormancy and germination.* Academic Press: San Diego, CA.
[19] Baskin CC and Baskin JM 2004 *Seed Sci. Res.* **14** 1–16 10.1079/SSR2003150
[20] Leymarie J *et al* 2012 *Plant Cell Physiol.* **53** 96–106 10.1093/pcp/pcr129
[21] Weitbrecht K, Müller K and Leubner-Metzger G 2011 *J. Ex. Bot.* **62** 3289–3309 10.1093/jxb/err030
[22] Basbouss-Serhal I, Leymarie J and Bailly C 2016 *J. Ex. Bot.* **67** 119–130 10.1093/jxb/erv439
[23] Jurdak R., Rodrigues G.d.A.G., Chaumont N., Schivre G., Bourbousse C., Barneche F., Bou Dagher Kharrat M. and Bailly C. (2022). *New Phytol*, 234: 850-866, https://doi.org/10.1111/nph.18038
[24] Buitink J. and Leprince O 20088 *C. R. Biol.* **331**, 10 10.1016/j.crvi.2008.08.002
[25] Ballesteros D, Pritchard HW and Walters C 2020 *Seed Sci. Res.* **30** 142-155 10.1017/S0960258520000239
[26] El-Maarouf-Bouteau H *et al* 2013 *Front. Plant Sci.* **4** 77 10.3389/fpls.2013.00077
[27] Bailly C 2019 *Biochem J.* **476**, 3019–3032 10.1042/BCJ20190159
[28] C. Bailly et al. 2008 *C R Biol.* **331**(10) 806-814 https://doi.org/10.1016/j.crvi.2008.07.022
[29] Alhamdan AM *et al* 2011 *American-Eurasian J. Agric. & Environ. Sci.*, **11**(3) 353-359
[30] Bazin J *et al* 2011 *J. Exp. Bot.*, **62**(2) 627-640 https://doi.org/10.1093/jxb/erq314
[31] Hay FR *et al.* 2022 *Front Plant Sci.* **13**, 891913 https://www.doi.org/10.3389/fpls.2022.891913
[32] Vertucci C W and Roos E 1993 *Seed Sci. Res.* **3** 201–213 10.1017/S096025850000179
[33] Walsh JL *et al* 2010 *J. Phys. D: Appl. Phys.* **43** 032001 JPhysD/43/03200
[34] Misra NN *et al* 2013 Am. Soc. Agr. Biol. Eng. **56** 1011-1016 10.13031/trans.56.9939
[35] Feizollahi E *et al* 2020 *Appl. Sci.* **10** 3530 10.3390/app10103530
[36] Machala Z *et al* 2007 *J. Mol. Spec.* **243** 194–201 10.1016/j.jms.2007.03.001
[37] Abdelaziz A, Ishijima T and Seto T 2018 *Phys. Plasmas* **25**, 043512 10.1063/1.5020271
[38] Yadav AK, Gautam CR and Singh P 2015 *J Mater Sci: Mater Elec* **26** 5001–5008 10.1007/s10854-015-3013
[39] L. Degutytė-Fomins et al. L, Pauzaite G, Zukiene R, Mildaziene V, Koga K, Shiratani M, 2020 *Jpn. J. Appl. Phys.* **59**, https://onlinelibrary.wiley.com/doi/pdfdirect/10.1002/pat.4764
[40] Walters C 1998 Seed Sci. Res. **8** 223-244 10.1017/S096025850000413X
[41] Buitink, J. & Leprince, O. Glass formation in plant anhydrobiotes: survival in the dry state. Cryobiology **48** 215–228. https://doi.org/10.1016/j.cryobiol.2004.02.011 (2004).
[42] Lehner A, Corbineau F and Bailly C 2006 *Plant Cell Physiol.* **47** 818-828 10.1093/pcp/pcj053
[43] Gerna D *et al* 2022 *J. Exp. Bot.* **73** 2631-2649 10.1093/jxb/erac024